\documentclass[letterpaper,10pt,journal,twocolumn]{IEEEtran}
\usepackage{amsmath,amsfonts,amssymb,bm}
\usepackage{graphicx,epsfig,psfrag,comment}
\usepackage{epstopdf}
\usepackage{multirow,slashbox}
\usepackage{algorithmic}
\usepackage[lined,boxed,commentsnumbered]{algorithm2e}
\usepackage[all]{xy}
\usepackage{varioref}%  smart page, figure, table, and equation referencing
\usepackage{wrapfig}%   wrap figures/tables in text (i.e., Di Vinci style)
\usepackage{threeparttable}% tables with footnotes
\usepackage{dcolumn,booktabs}%   decimal-aligned tabular math columns
\newcolumntype{d}{D{.}{.}{-1}}
\usepackage{nomencl}%   nomenclature generation via makeindex
\makeglossary
\usepackage{subfig}% subcaptions for subfigures
\usepackage{comment,cite}
\usepackage{remark}
\usepackage{xfrac}
\usepackage{mathrsfs}
\usepackage{fixltx2e}
\usepackage{xcolor}

\allowdisplaybreaks

\newcommand\numberthis{\addtocounter{equation}{1}\tag{\theequation}}

\newtheorem{problem}{Problem}

\newremark{remark}{Remark}
\newremark{proof}{Proof}

 \SetAlCapSkip{1em}
\SetKwInOut{Input}{Input}\SetKwInOut{Output}{Output}
\renewcommand{\baselinestretch}{0.94}

\title{\textbf{Min-Max Q-Learning for Multi-Player Pursuit-Evasion Games}}

\author{Jhanani Selvakumar \and  Efstathios Bakolas
\thanks{J. Selvakumar
is a PhD candidate at the Department of Aerospace Engineering
and Engineering Mechanics, The University of Texas at Austin,
Austin, TX 78712-1221, USA, Email: jhanani@utexas.edu}
\thanks{E. Bakolas
is an Associate Professor at the Department of Aerospace Engineering
and Engineering Mechanics, The University of Texas at Austin,
Austin, TX 78712-1221, USA, Email: bakolas@austin.utexas.edu}}

\begin{document}

\maketitle

\begin{abstract}
In this paper, we address a pursuit-evasion game involving multiple
players by utilizing tools and techniques from reinforcement
learning and matrix game theory. In particular, we consider the
problem of steering an evader to a goal destination while avoiding
capture by multiple pursuers, which is a high-dimensional and
computationally intractable problem in general. In our proposed
approach, we first formulate the multi-agent pursuit-evasion game as
a sequence of discrete matrix games. Next, in order to simplify the
solution process, we transform the high-dimensional state space into
a low-dimensional manifold and the continuous action space into a
feature-based space, which is a discrete abstraction of the original
space. Based on these transformed state and action spaces, we
subsequently employ min-max Q-learning, to generate the entries of
the payoff matrix of the game, and subsequently obtain the optimal
action for the evader at each stage. Finally, we present extensive
numerical simulations to evaluate the performance of the proposed
learning-based evading strategy in terms of the evader's ability to
reach the desired target location without being captured, as well as
computational efficiency.
\end{abstract}


\section{Introduction}
Interaction between multiple agents is present in nature (e.g.
predation) as well as human-built constructs (such as markets,
traffic situations, and defense). In order to better understand and
predict the potential outcomes of such interactions among multiple
agents, each with their own aims, preferences and resources, we have
to understand their underlying decision-making mechanisms. Uncertain
interactions among agents can be studied within the framework of
dynamic non-zero-sum multi-player games. A special class of such
problems are pursuit-evasion games (PEGs) with multiple players
which seek to capture or evade each other. In this paper, we
consider the problem of steering an evader to a target location
while avoiding capture by multiple pursuers. Finding the exact
solution to such a pursuit-evasion game can be a complex task due to
the high dimensionality of the problem. In particular, the state
space of the PEG consists of high-dimensional vectors that
correspond to the concatenations of the position and velocity
vectors of all the players involved. Similarly, the combined action
space of all the players can be also high-dimensional.
%These state and action spaces
%can be continuous or discrete.

In this paper, the state space and the action space of the PEG are
both continuous. Obtaining exact solutions to this class of problems
can be a very complex task and thus, the characterization of
approximate solutions is preferred in practice. We use min-max
Q-learning to obtain an approximation of a Q-function that can
characterize the evader's payoff (reward) for actions taken by the
different players from any state. It is desirable to have the
results of the learning process be independent of specific
parameters of the training instances, particularly in terms of the
number of players and their speeds. To this aim, the learning is
performed in a low-dimensional manifold, which is obtained by
applying a nonlinear transformation to the continuous state space,
and a discrete abstraction of the continuous action space
(feature-based action space). In the learning space, the interaction
of the players in the pursuit-evasion game is described purely in
terms of times to capture under different conditions, and the
discrete actions are designed to correspond to the intent of the
different players. The approximate Q-function is used to construct a
matrix that describes a two-player game per stage. The solution to
the matrix game yields the evader's action strategy at the current
stage. The pursuers may either employ the strategy induced by the
latter game or a predetermined policy such as relay pursuit (i.e.,
only the nearest pursuer to the evader tries to capture the latter).
%whereas the pursuers may employ the previously decided strategy of
%relay pursuit or base their decision on the Q-function.
%%

\noindent \textit{Literature survey:} Multi-player pursuit evasion
games are extensively used to model interactions in economics,
biology and defense, to name a few \cite{c1,c2,c3}. Multi-player
games are typically non-zero sum, and may be played in continuous or
discrete time. Non-cooperative multi-player games, including matrix
games, are discussed in detail by Basar and Olsder \cite{c4}. Matrix
games, also discussed by Zaccour et. al in \cite{c5}, could be
single act (static games) or multi-act (repeated or dynamic games)
\cite{c6}. Well-known results pertaining to existence of Nash
equilibria for single-stage matrix games can be found in~\cite{c4}.
The question of existence of Nash equilibria in repeated games is
addressed by the so-called folk theorem~\cite{c7}. Multi-player
pursuit-evasion games with the emphasis placed either on capture or
evasion have been addressed in
\cite{c8,c9,c10,c11,c12,c13,c15,garcia2017geometric,p:selvakumarCYB2019}.
The latter references use approaches which are based on geometric
arguments, such as dynamic Voronoi partitions, switching strategies,
Voronoi-based roadmaps as well as greedy pursuit policies.

A Markov game-formulation of the discrete multi-agent PEG in an
uncertain environment is treated using a matrix game approach in
\cite{c16}. In our previous work \cite{c14}, the multi-agent dynamic
non-zero-sum game was re-formulated as a sequence of discrete
zero-sum matrix games. The evader's myopic strategy corresponded to
the solution to a linear program that was solved at each stage. In
this paper, we retain this basic solution structure, however, we
improve our previous formulation of the matrix game using Q-learning
to learn the elements of the payoff/cost matrix. Q-learning
techniques are popular tools for the characterization of optimal
policies in decision making problems in uncertain environments
involving either a single player or multiple
players~\cite{sutbarto,c17,c18,c19,c20,bilgin2015approach}. For
two-player zero-sum games that are cast as Markov games, min-max
Q-learning is applied where the definition of the Q-function is
modified to suit the presence of a second, independent
decision-maker. The use of min-max Q-learning has been demonstrated
for simple problems with discrete state and action spaces in
\cite{c17} and an integral Q-learning algorithm for continuous
differential games is presented in \cite{c20}. An exposition of
different Q-learning algorithms for Markov games is found in
\cite{c19}, where the authors also present an improved learning
algorithm to compute mixed policies. Multi-agent reinforcement
learning, using different value functions for the agents, is
presented in \cite{hu2015multiagent}. In \cite{bilgin2015approach},
concurrent Q-learning is used by both pursuing and evading agents to
learn their optimal policies. A critical overview of reinforcement
learning for multi-agent problems can be found in
\cite{critsurv,c23}. Finally, examples of the use of deep learning
to solve multi-agent problems in dynamic environments are presented
in \cite{sui2018path,wang2018learning}.

\noindent \textit{Contributions:} In this paper, we first propose a
systematic way to reduce a multi-agent dynamic non-zero-sum game to
a sequence of two-player static zero-sum games each of which can be
formulated as a linear program. In this set-up, the payoffs of each
static game will determine the action of the evader at that stage,
and this, in turn, will determine the performance of the evader's
strategy. We construct the payoffs taking into account (i) a risk
metric for the evader with regard to its capture, and (ii) the time
required by the evader to reach the target location. In order to
refine the performance of the evader's strategy, we employ a min-max
Q-learning algorithm to determine the entries of the payoff matrix
at each stage of the game.

In our approach, learning takes place in a low-dimensional nonlinear
manifold (learning space) embedded in the original high-dimensional
state space. The states in this reduced space capture the salient
features of the pursuit-evasion game using time-of-capture
parameters rather than position and velocity. Consequently, the new
state space is invariant with respect to the number of players or
the dynamic model of the players which is a key property of our
proposed approach. Furthermore, we associate the continuous action
space of the players to a feature-based action space which is a
discrete space comprised of a small number of actions. The set of
discrete actions is informed by the members of the continuous action
space. These actions affect the time-to-capture, and/or the
time-to-target, of one or more players. Our choice of discrete
actions does not restrict the movement of the evader to
pre-specified spatial directions, but rather, allows the evader to
move in different spatial directions as required by the
configuration of each game.

\noindent \textit{Structure of the paper:} In Section II, we define
the target-seeking evasion problem, and in Section III, we formulate
the corresponding multi-act two-person zero-sum game. The set-up for
min-max Q-learning in the context of our problem, and the learning process are described in
Section IV and Section V respectively. Numerical
simulations and observations are given in Section VI. In Section
VII, we present concluding remarks.

\section{Formulation of target-seeking evasion problem}\label{pr:form}

Consider a pursuit-evasion game with $N$ pursuers and one evader
that takes place in a domain $\mathcal{D} \subseteq \mathbb{R}^{2}$.
We will assume that there is an upper bound on the duration of the
game which is known a priori and is denoted by $\bar{T}_f>0$. For
the sake of simplicity, we demonstrate the proposed solution
technique in a reach-avoid game where all the players have single
integrator dynamics. At any given time $t \in [0,\bar{T}_f]$, the
state (position) of the $i^{th}$ pursuer $P_i$, where $i \in
\mathcal{I}:=\{1, \dots, N\}$, is denoted by $\bm{x}_{ic} \in
\mathbb{R}^2$, whereas the state (position) of the single evader $E$
at time $t$ is denoted by $\bm{x}_{ec} \in \mathbb{R}^2$. The
dynamic equations are given by:
\begin{subequations}\label{cdyn}
\begin{align}
{\dot{\bm{x}}}_{ic}(t)&=v_{p_i}\bm{u}_{ic}(t) ,~~~~\bm{x}_{ic}(0)=\bar{\bm{x}}_i, \\
{\dot{\bm{x}}}_{ec}(t)&=v_{e}\bm{u}_{ec}(t)
,~~~~\bm{x}_{ec}(0)=\bar{\bm{x}}_e,
\end{align}
\end{subequations}
\noindent where $v_{p_i}>0$ and $v_e>0$ denote respectively the
maximum speeds of the $i^{th}$ pursuer and the evader. Further,
$\bm{u}_{ic}(t) \in \mathbb{R}^2$ and $\bm{u}_{ec}(t) \in
\mathbb{R}^2$ denote the inputs of the $i^{th}$ pursuer and the
evader at time $t$, respectively, and are assumed to take values in
the set $\mathcal{U}:=\{\bm{u} \in \mathbb{R}^2: \|\bm{u}\| = 1
\textrm{ or }\bm{u} = \bm{0} \}$. The zero control input will be
used by the players only when the game terminates (either by capture
of the evader by at least one pursuer, or by the evader reaching its
target location).

The framework of matrix games applied to a continuous-time dynamic
game requires the choice of actions, repeatedly, at each instant of
time. While the state space of the game and the action space are
both continuous, we will consider a discrete-time state-space model
for the players in Eq. \eqref{cdyn}, wherein the control input is
piece-wise constant. That is, the control input is a constant vector
in each time interval $[k\Delta t, (k+1) \Delta t)$, where $k$ is a
positive integer and $\Delta t \ge 0$ is the sampling period. The
game consists of $\bar{K}+1$ finite stages at most, with a constant time step $\Delta t
>0$, such that $\bar{T}_f := \bar{K} \Delta t$. Note that at time $t
= k \Delta t$, the discrete-time state of the $i^{th}$ pursuer, is
denoted by ${\bm{x}}_{id}(k)$, and is defined as ${\bm{x}}_{id}(k):=
 \bm{x}_{ic}(k \Delta t)$. Similarly, the
discrete-time states are defined for the other players. If $k \in
\{0,...,\bar{K}\}$ denotes the current stage of the game, and
$\bm{x}_{id}(k)$ is the state vector of the $i^{th}$ pursuer at that
stage and $\bm{x}_{ed}(k)$ is the state vector of the evader, the
discrete-time state-space model for the players is described by the
following equations:
\begin{subequations}\label{ddyn}
\begin{align}
{\bm{x}}_{id}(k+1)&=\bm{x}_{id}(k) + (v_{p_i} \Delta t) \bm{u}_{id}(k),~~\bm{x}_{id}(0)=\bar{\bm{x}}_i, \\
{\bm{x}_{ed}}(k+1)&=\bm{x}_{ed}(k) + (v_e \Delta t)
\bm{u}_{ed}(k),~~\bm{x}_{ed}(0)=\bar{\bm{x}}_e,
\end{align}
\end{subequations}
\noindent where $\bm{u}_{id}(k)$, $\bm{u}_{ed}(k) \in \mathcal{U}$
denote the inputs of the $i^{th}$ pursuer and the evader at stage
$k$, respectively. The discrete-time control input corresponds to a
piece-wise constant continuous-time signal, that is,
$\bm{u}_{ic}(\tau):=\bm{u}_{id}(k)$ for all $\tau \in [k \Delta t,
(k+1) \Delta t)$ for all $k \in \{0,...,\bar{K}-1\}$.

In this paper, we assume that the pursuers have a preferred
strategy. In particular, they engage in what is known as ``relay
pursuit'' with corresponding \textit{relay metric} the minimum
time-to-capture~\cite{c11}. At each instant of time, the active
pursuer is the one corresponding to the smallest time-to-capture
(among the pursuers). In relay pursuit, only one pursuer is active,
that is, engages in pursuit, whereas the other pursuers are
stationary. The assignment of the active pursuer can change with
time based on which pursuer has the least time-to-capture. If the
active pursuer is designated by the index $i^{\star}$, then, for all
$t \in [0,t_f]$, the feedback strategy of the $i^{th}$ pursuer is
\begin{equation}
    \bm{u}_{id}^*(\bm{x}_{ed}, \bm{x}_{id})=
    \begin{cases}
     \bm{r}_{cd}/ \|\bm{r}_{cd}\|, & \mathrm{if} ~ i= i^{\star},\\
     \bm{0}    & \mathrm{otherwise},
    \end{cases}
    \label{purstrat}
  \end{equation}
where $\bm{r}_{id}:=\bm{x}_{ed}-\bm{x}_{id}$ is the relative
position vector of the evader with respect to the $i^{th}$ pursuer.
Relay pursuit is a suitable choice of strategy for a group of
pursuers in which each agent wishes to remain spatially localized,
or to conserve resources.

Capture occurs when the evader's distance from at least one pursuer
is less than the radius of capture $\ell>0$. That is, the game will
terminate in capture at stage $K \in \{0, \dots,\bar{K}\}$ (for a
given positive integer $\bar{K}$), if there exists $i \in
\mathcal{I}: \|\bm{x}_{id}(K)-\bm{x}_{ed}(K)\| \le \ell$. The target
is denoted by $\bm{x}_T\in \mathbb{R}^2$. The evader is considered
to be successful in reaching the target if
$\|\bm{x}_{ed}(K)-\bm{x}_T\| \le \epsilon$, where $\epsilon>0$. The
multi-player reach-avoid problem in discrete-time is stated as
follows:

\begin{problem} \label{raprob_disc}
\textit{Let $\bar{K}$ be a positive integer and let us assume that
the initial positions of the pursuers $\bar{\bm{x}}_i$, for $i \in
\mathcal{I}$, and the initial position of the evader,
$\bar{\bm{x}}_e$, be given. Find a sequence of control inputs $\{
\bm{u}_{ed}(k)\}_{k=0}^{K-1}$, where $K$ is a positive integer with
$K \leq \bar{K}$ ($K$ corresponds to the free terminal stage), which
will guide the evader to the target
 $\bm{x}_T \in \mathcal{D}$ within the desired tolerance, while avoiding capture,
that is, $\|\bm{x}_{ed}(K)-\bm{x}_T\| \le \epsilon$, and
$\|\bm{x}_{ed}(k)-\bm{x}_{id}(k)\| > \ell, ~\forall i \in
\mathcal{I}, \forall k \in \{0,1,..,{K}\}$. }
\end{problem}

\noindent We assume that the players have perfect information about the states of all the
players of the game at all times. In addition, the target $\bm{x}_T$
is known only to the evader. Note that the evader has two goals: (a)
reaching the target location, and (b) avoiding capture, whereas the
group of pursuers has only one: to achieve capture of the evader as
soon as possible.

For compactness of notation, subsequently, the subscript $d$ is dropped
from the state and control vector notations at the corresponding stage $k$.

\section{Matrix game formulation of the multi-player PEG} \label{mat:form1}
In this section, we re-formulate the discrete-time non-zero-sum game
with $N+1$ players described in Problem \ref{raprob_disc} as a
multi-act two person zero-sum matrix game. The group of pursuers is considered as a single entity, $P$, with the ability to deploy exactly one of the pursuers at a given instant of time. At each stage $k$ of the
latter game, we consider a matrix $M_k \in \mathbb{R}^{N \times
(N+1)}_{\ge0}$, whose entries are the payoffs to $E$ at that stage.
Each row of $M_k$ represents a pure strategy played by $P$ and each
column, a pure strategy played by $E$.
The decision space available to the players (the space of control
inputs for $P$ and $E$) is continuous and contains infinite number
of actions. We consider a restricted decision space for $P$,
including only the actions that appear ``integral'' to the pursuers'
goal of capturing the evader. In particular, $P$ has exactly $N$
pure strategies, where the $i^{th}$ pure strategy corresponds to the
case where only the $i^{th}$ pursuer goes after the evader.
Similarly, $E$'s restricted decision space consists of $N+1$
actions, where the first $N$ correspond to evasion from each pursuer
in turn (the $j^{th}$ action is to avoid only the $j^{th}$ pursuer),
and the $(N+1)^{th}$ action is the target-seeking behavior, which
means that the evader directly heads towards the target.

\noindent Let $i$ be the row index of $M_k$ and $j$ be the column
index, where $i \in \mathcal{I}$ and $j \in \mathcal{J}:=\{1, \dots,
N+1\}$. If we consider the first $N$ columns of $M_k$, each entry
$M_k(i,j)$ with $i=j$ is the payoff for the two-player zero-sum game
between only the $i^{th}$  pursuer and the evader. Every other entry
$M_k(i,j)$, with $i \neq j,$ represents a case when $E$ tries to
evade from the $j^{th}$ pursuer, when actually the $i^{th}$ pursuer
is active. This situation can happen because while the $E$ knows the
states of all the pursuers, it does not know the action chosen by
$P$ at the same stage. Finally, the last column of the matrix $M_k$
represents cases where the evader is directly headed towards the
target $\bm{x}_T$, and only one pursuer is active per row. A
schematic construction of the matrix $M_k$ is shown in Table
\ref{stattab}.

\begin{table}
\caption{Entries in the payoff matrix $M_k$, for the case of $N=2$.}
\begin{center}
    \begin{tabular}{ | l | l | l |}
    \hline
    (1,1)  & (1,2) & (1,3)\\
    $P_1$ in pursuit & $P_1$ in pursuit &  $P_1$ in pursuit \\
  $E$ evading $P_1$ & $E$ evading $P_2$ & $E$ seeks $\bm{x}_T$\\\hline
    (2,1) & (2,2) & (2,3)\\
    $P_2$ in pursuit & $P_2$ in pursuit  & $P_2$ in pursuit \\
    $E$ evading $P_1$ &$E$ evading $P2$ & $E$ seeks $\bm{x}_T$\\\hline
    \end{tabular}
\end{center}
\label{stattab}
\end{table}

The payoff matrix of the matrix game has to be updated at every
stage of the game because the players move between every pair of
successive stages and the entries of the payoff matrix must change
to reflect the underlying dynamic evolution of the game. For this
reason, our matrix game formulation differs from the class of games
known as repeated static games. Constructing and updating the
entries of the payoff matrix is central to developing a successful
strategy for the evader that can lead it to the target while
avoiding capture. The values in the payoff matrix must represent the
two-fold goal of the evader.

\subsubsection{Time metrics of the reach-avoid game}
Next, we introduce a number of time-metrics that will be used in the
subsequent analysis. All these metrics are associated with the
minimum positive solution to the following equation~\cite{c6}:
\begin{equation}
(v_e^2-v_{p_i}^2) \phi^2+ 2 (\langle \bm{r}_i, v_e \bm{u}_e \rangle
- \ell v_{p_i}) \phi + \langle \bm{r}_i, \bm{r}_i \rangle -\ell^2
=0, \label{toc}
\end{equation}
where $\bm{r}_i:=\bm{x}_e-\bm{x}_i$ and $\bm{x}_e$ and $\bm{x}_i$
correspond to the position of the evader and the $i^{th}$ pursuer at
the beginning of the game (at stage $k$, where $0\leq k < \bar{K}$),
for different values of the vector $\bm{u}_e$. In particular, the
min-max time-to-capture of the evader by the $i^{th}$ pursuer,
ignoring all other players, is denoted as $\phi_c(\bm{x}_e,
\bm{x}_i)$ and is defined as the minimum positive real solution to
the equation~\eqref{toc} when $\bm{u}_{e}:=\bm{r}_i/ \|\bm{r}_i\|$
(the $i^{th}$ pursuer uses the control input $\bm{u}_{i}:=\bm{r}_i/
\|\bm{r}_i\|$). Similarly, let $\phi_a(\bm{x}_e, \bm{x}_i,
\bm{u}_e)$ denote the minimum time-to-capture of the evader by the
$i^{th}$ pursuer when the evader is moving in a randomly chosen
fixed direction denoted by $\bm{u}_e$. Ideally, this particular
maneuver does not avoid any pursuer in particular, but may delay
capture by exploiting the relative positioning of the pursuers
around the evader. It is possible to improve this particular evasion
maneuver by taking into account the dynamic models of the players.
In particular, $\phi_a(\bm{x}_e, \bm{x}_i, \bm{u}_e)$ is the minimum
real positive solution to Eq. \eqref{toc}, where $\bm{u}_e$
represents the direction of the evader's random maneuver.

Finally, let $\phi_s(\bm{x}_e, \bm{x}_i, \bm{x}_T)$ denote the
minimum time-to-capture of the evader by the $i^{th}$ pursuer when
the evader is directly headed towards the target location, in which
case, $\phi_s(\bm{x}_e, \bm{x}_i, \bm{x}_T)$ is the minimum real
positive solution to Eq. (\ref{toc}) with
$\bm{u}_e=({\bm{x}_T-\bm{x}_e})/{\|\bm{x}_T-\bm{x}_e\|}$. Note that
in this case, the evader does not maneuver to avoid the pursuer, and
only the pursuer is maneuvering to minimize the capture time.
Finally, let $\phi_T(\bm{x}_e,\bm{x}_T)$ denote the minimum time for
the evader to reach the target.

\subsection{Elements of the payoff matrix} \label{elem}
In the current matrix game formulation, each element of the payoff
matrix is a numerical value that reflects the two-fold objective of
the evader: (a) to avoid capture and (b) to reach the target
location $\bm{x}_T$. The two components of each entry are the time
that $P$ would take to capture $E$, and the extent to which $E$'s
heading is towards $\bm{x}_T$ from its current location. The
target-seeking component of $E$'s velocity is equal to $\cos
\theta$, where $\theta$ is the angle between the vectors $\bm{u}_e$
and $\bm{x}_T-\bm{x}_e$.

Because at least two players have moved in the time between the
current stage and the previous stage, one has to re-construct the
payoff matrix at each stage. While a future payoff is as important
as the present payoff for consideration, it is difficult to estimate
the payoff that the evader will receive at the end of $K$ stages,
since the payoffs at each stage are dependent on the players' states
in the current stage. Thus, the history of moves in previous play is
reflected in the changing payoff values, although this information
is not available directly to the players as a strategy recall.

The payoffs are designed to reflect the long-term effects of each
action. In this case, since we have an upper bound $\bar{T}_f$ on
the duration of the game, the time-of-capture component is bounded.
For each pair of pure strategies $(i,j)$, we calculate the minimum
time-of-capture of $E$ by the pursuing agent $i$. Note that $E$ will
play the strategy corresponding to evasion from the pursuer $j$. If
capture is not possible, we set the value to $\bar{T}_f$. The
target-seeking component is given by $\cos \theta$, as described
earlier. Algorithm \ref{algo_cost2} shows the main steps for the
assignment of payoffs to $M_k$:
\begin{algorithm}
\caption{Payoff Assignment to $M_k$}
\SetKwInOut{Input}{input}\SetKwInOut{Output}{output}
\Input{$\bm{x}_e$, $\bm{x}_T$, $\bm{x}_i ~\forall i \in \mathcal{I}$, $k$, $\bar{T}_f$}
\Output{$M_k$}
\BlankLine

\For{$i\leftarrow 1$ \KwTo $N$} {
        \For{$j\leftarrow 1$ \KwTo $N+1$} {

                   $T_c=\min(\phi_a(\bm{x}_e, \bm{x}_i, \bm{u}_j), \bar{T}_f)$\\
                   $G_c=\frac{\langle \bm{u}_e, \bm{x}_T-\bm{x}_e \rangle}{\|\bm{u}_e\|\|\bm{x}_T-\bm{x}_e \|}$\\

            $M_{k1}(i,j)=T_c$\\
                $M_{k2}(i,j)=G_c$
    }
}
$\hat{M}_{k1}=\frac{M_{k1}}{\max_{i,j}{M(i,j)_{k1}}}$\\
$M_{k}=\hat{M}_{k1}+M_{k2}$

\label{algo_cost2}
\end{algorithm}

\noindent Note that we normalize the evasion component which is
given by the matrix $M_{k1}$ using the maximum entry of the matrix.
This ensures that all evasion components have values between zero
and unity, with magnitude similar to the target-seeking component.
The summation of the two components in this manner is a standard
practice in multi-objective optimization where the objectives are
combined into one global criterion \cite{aroramult}.

\subsection{Solution to the matrix game}
We proceed to solve for the equilibrium strategies for the players.
An equivalent non-zero sum formulation for our problem would
consider the whole $N+1$ player game, with cost assignments that are
functions of the states of all players. Consequently, verifying the
existence of an equilibrium set of pure strategies is a hard
problem, in the sense that it would require an exhaustive search
among all possibilities. However, we know that a two-player zero-sum
multi-act game which is finite, admits a saddle point solution in
mixed strategies\cite{c4}.

The min-max solution to the matrix game at stage $k$ consists to two
vectors of probabilities (discrete probability distributions), one
for the evader $E$ which is denoted by $\bm{\pi}_{ek}$, and one for
the group of pursuers as a whole which is denoted by
$\bm{\pi}_{pk}$. These vectors correspond to the \textit{mixed
strategies} of the two players. Each entry of the vector
representing the mixed strategy of a player corresponds to the
probability with which each pure strategy is to be chosen, when the
game is played with infinite turns. The discrete action chosen at
each stage $k$ is a random sample from the probability distribution
assigned by the mixed strategy. Note that the matrix formulation
enables the conversion of the $(N+1)$-player dynamic game to the
two-player decision-making problem with discrete action choices for
each player.

In particular, at stage $k$, the mixed strategy for $P$,
$\bm{\pi}_{pk} \in \mathbb{R}^N$, given by
$\bm{\pi}_{pk}:=\left[\pi_{pk}^1, \dots , \pi_{pk}^{N}
\right]^{\mathrm{T}}$, with $\sum_{i=1}^{N} {\pi}^i_{pk}=1$, and
${\pi}_{pk}^i \ge 0$. In other words, the mixed strategy for the $N$
pursuers can be defined, by considering a member of the simplex of
dimension $N-1$, for the first $N-1$ probabilities (the $N^{th}$
probability is obtained as the difference of their sum from unity).
Similarly for $E$, the mixed strategy is denoted by $\bm{\pi}_{ek}
\in \mathbb{R}^{N+1}$, where $\bm{\pi}_{ek}:=\left[\pi_{ek}^1, \dots
, \pi_{ek}^{N+1} \right]^{\mathrm{T}}$, and $\sum_{j=1}^{N+1}
{\pi}_{ek}^j=1$, and ${\pi}_{ek}^j \ge 0$.  The $i^{th}$
(respectively, $j^{th})$ entry of the vector $\bm{\pi}_{pk}$
(respectively, $\bm{\pi}_{ek}$) represents the probability of the
$i^{th}$ (respectively, the $j^{th})$ pure strategy being employed
by $P$ (respectively, $E$). We only choose one pure strategy for
each player per stage of the game. However, in the limiting case
(for a game of infinite stages), the frequency at which each pure
strategy is chosen converges to the probabilities assigned to the
pure strategies by the mixed strategies $\bm{\pi}_{pk}$ and
$\bm{\pi}_{ek}$. The problems of computing the vectors
$\bm{\pi}_{pk}^*$ and $\bm{\pi}_{ek}^*$ that correspond to the
saddle point of the (min-max) game can be formulated as Linear
Programming (LP) problems \cite{c5}, which can be solved using
readily available solvers. In particular, $\bm{\pi}^{\star}_{pk}$
corresponds to the solution to the following LP problem:
\begin{align*}
\min_{(p,\bm{\pi}_{pk})} ~~~&p\\
~~\textrm{subject to}~~~p \mathbf{1} &\ge M_k^{T} \bm{\pi}_{pk},\\
\bm{\pi}_{pk} &\ge 0,\\
\mathbf{1}^T \bm{\pi}_{pk}&=1.
\numberthis
\label{LP1}
\end{align*}
where $\mathbf{1}:=\left[1,\dots,1 \right]^\mathrm{T} \in
\mathbb{R}^N$. Similarly, for the computation of
$\bm{\pi}^{\star}_{ek}$, one has to solve the following LP problem:
\begin{align*}
\max_{ (q, \bm{\pi}_{ek})} ~~~&q\\
~~\textrm{subject to}~~~q \mathbf{1} &\le M_k \bm{\pi}_{ek},\\
\bm{\pi}_{ek} &\ge 0,\\
\mathbf{1}^T \bm{\pi}_{ek}&=1.
\numberthis
\label{LP2}
\end{align*}
Note that the LP problem~\eqref{LP1} whose solution is
$\bm{\pi}^{\star}_{ek}$ is the dual to the LP problem~\eqref{LP2}
whose solution is $\bm{\pi}^{\star}_{pk}$. From strong duality, we
have that $\max_{\bm{\pi}_{ek}} \bm{\pi}_{pk}^{*T} M_k
\bm{\pi}_{ek}=\min_{\bm{\pi}_{pk}}
\bm{\pi}_{pk}^T M_k \bm{\pi}_{ek}^*$. %where $\bm{\pi}_{pk}^*$ and
%$\bm{\pi}_{ek}^*$ are the mixed strategies computed.

The two LPs (\ref{LP1}) and (\ref{LP2}) are solved at each stage of the game after we recompute
$M_k$ to account for the change of the players' locations between
successive stages. In practice, the actions (pure strategies) for a
particular stage of the game are obtained as random samples from the
discrete distributions given by $\bm{\pi}_{pk}^*$ and
$\bm{\pi}_{ek}^*$ for that stage of the game.

\section{New matrix formulation of the multi-agent PEG with Q-learning}
In the formulation of the matrix game described so far, the size of
the matrix depends on $N$, the number of pursuers. In addition, the
action space of the evader is comprised of actions that restrict the
evader to switch between pure evasion (try to delay as much as
possible the capture by the active pursuer) and goal-seeking (reach
the goal destination as fast as possible). Next, we propose a new
payoff matrix, whose structure renders it independent of the number
of pursuers, as well as the number of actions that are available to
the evader. Further, we are interested in improving the payoff
function that determines the payoff matrix, so that the evader's
strategy performs better in terms of guiding the evader to the
target without getting captured. To construct this new payoff
function, we propose to use Q-learning techniques, with the
intention of recognizing possible patterns of play in the
multi-agent game, which lead to a better evasion strategy.

At each stage $k$ of the game, we consider a new matrix $M_k \in
\mathbb{R}^{N_p \times N_e}_{\ge 0}$, whose entries are the payoffs
(rewards) to $E$ at that stage. Alternatively, the entries of $M_k$
can be viewed as the costs incurred by the group of pursuers as a
whole. Again, since the payoffs to the evader and the pursuer group
sum to zero, the matrix represents a zero-sum game with the evader
as the maximizing agent and the group of pursuers as the minimizers.
Each row of $M_k$ represents a pure strategy, $p_i$, $i\in
\{1,...,N_p\}$, played by the group of pursuers and each column
represents a pure strategy, $e_j$, $i \in \{1,..,N_e\}$, which is
played by $E$. Each pure strategy qualitatively describes an action
taken by the corresponding player or players. Note that the learning
space which is used in the learning process is the output of a
nonlinear transformation applied to the original (high-dimensional)
state space. Further, the action space used for learning corresponds
to a feature-based action set, which is an abstraction of the
physical action set available to the players.

\subsection{Discrete action sets for the players}

\begin{figure*}[!htb]
\centering %
\begin{tabular}{cccc}
 \subfloat[Action $e_1$ for $N=3$]{\label{a1}\includegraphics[width=0.22\textwidth]{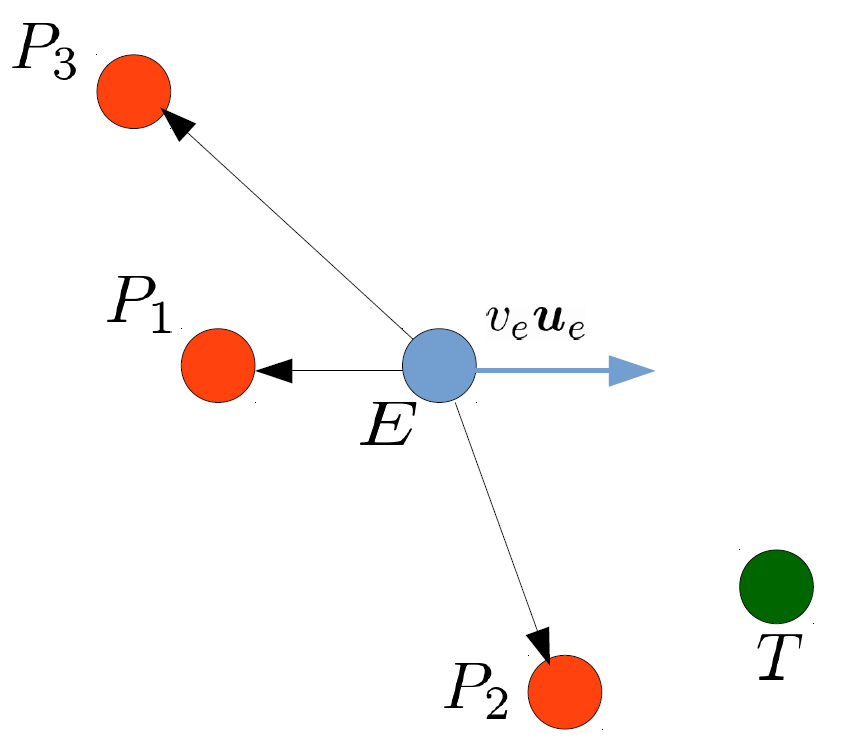}}& %0.75\linewidth
 \subfloat[Action $e_2$ for $N=4$]{\label{a2}\includegraphics[width=0.22\textwidth]{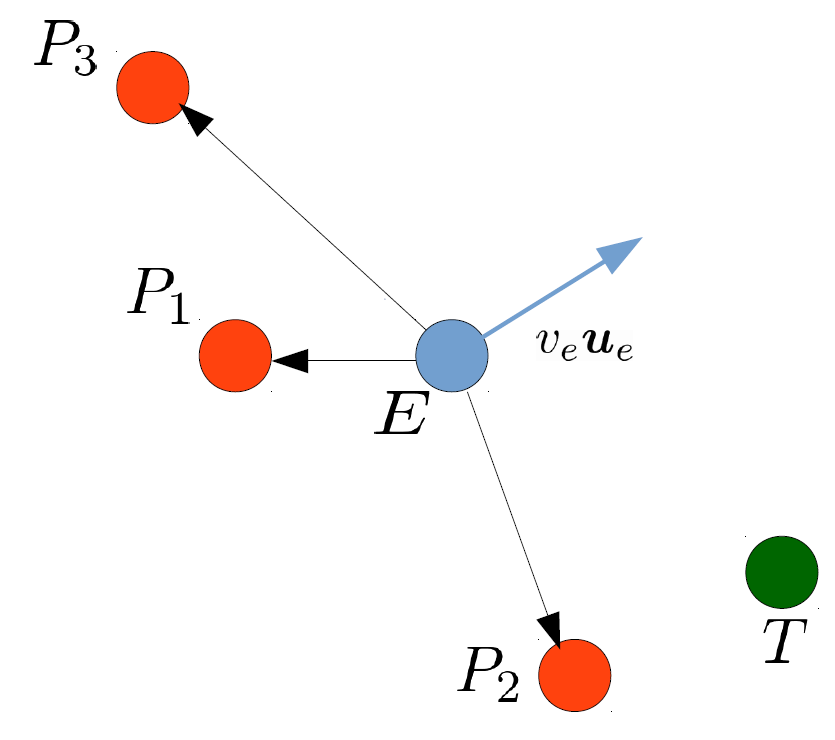}}&
 \subfloat[Action $e_3$ for $N=3$]{\label{a3}\includegraphics[width=0.22\textwidth]{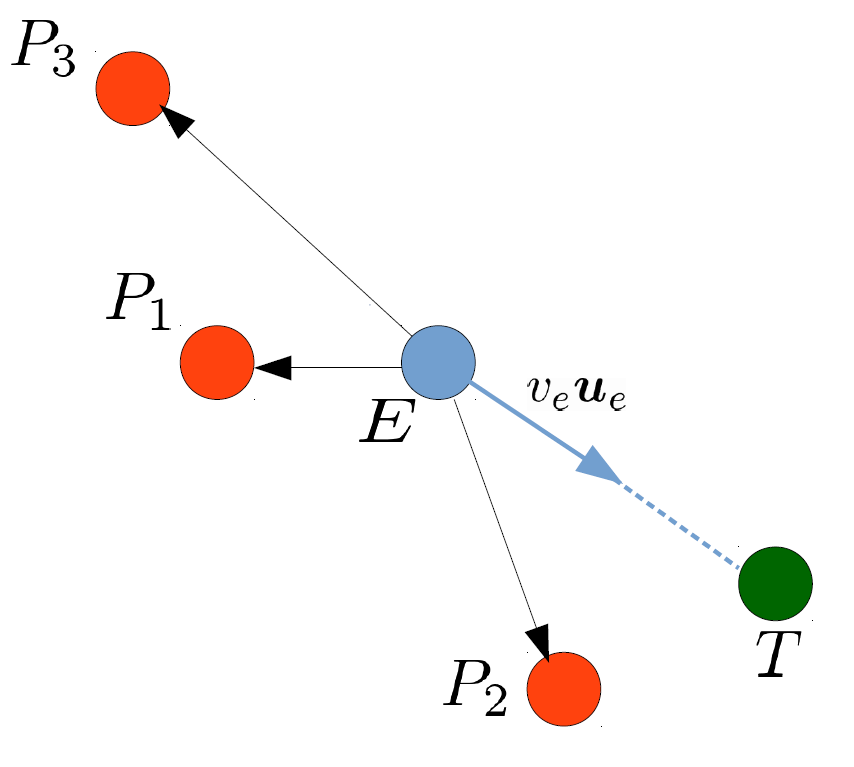}}& %0.75\linewidth
 \subfloat[Action $e_4$ for $N=4$]{\label{a4}\includegraphics[width=0.22\textwidth]{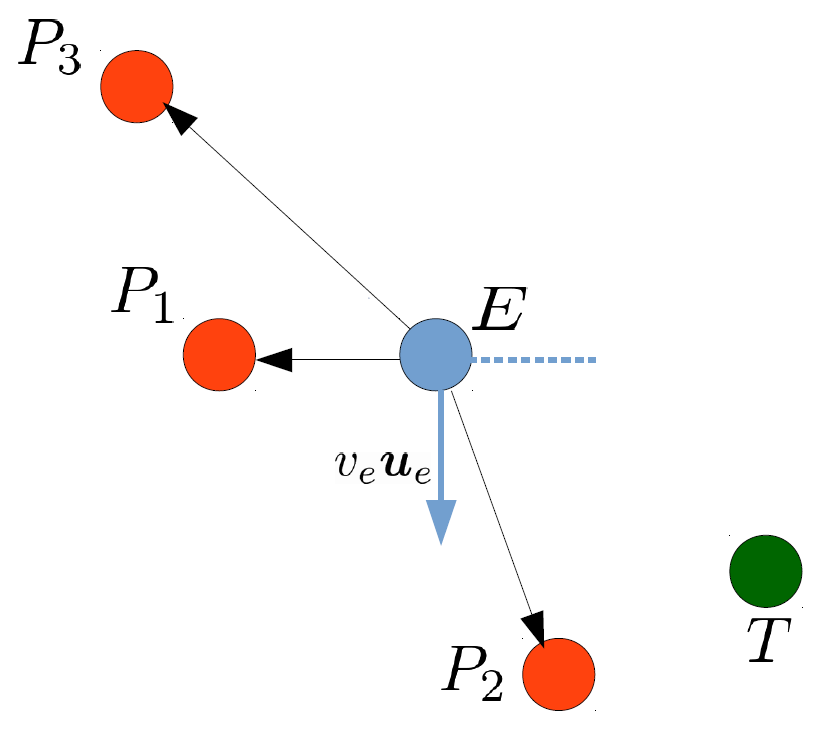}}\\  %0.75\linewidth
\end{tabular}
\caption{\small{Schematic physical realization of evasive actions
for an evader with simple dynamics. The pursuers are represented as
red circles and the evader in blue. The blue arrow indicates the
direction of motion of the evader determined by the corresponding
evasion action.} }\label{evacts} \vspace{0.6cm}
\end{figure*}

The set of actions for the evader, which is denoted by
$\mathcal{A}_e$, is a finite set comprised of $N_e$ elements.
Similarly, a finite set $\mathcal{A}_p$ comprising $N_p$ actions is
considered for the group of pursuers. To streamline the subsequent
discussion and analysis, we choose $N_e=4$ and $N_p=2$. The choice
of actions for each player (either as an individual or as a group)
is motivated by the qualitative effect of each action on the game.
The actions that are considered for the group of pursuers are
denoted by $p_i \in \mathcal{A}_p, i \in \{1,...,N_p\}$, and are
described qualitatively as follows:
\begin{enumerate}
\item $p_1$: the pursuers engage in relay pursuit, that is, only the pursuer
closest to the evader engages in pursuit while the others maintain their current state of operation (remaining static).
\item $p_2$: all pursuers actively engage in pursuit of the evader simultaneously and non-cooperatively.
\end{enumerate}
Note that the focus of this work is on the characterization of an
evasion strategy rather than a pursuit strategy. For this reason, we
have assumed that the pursuers have a preferred strategy, namely
relay pursuit \cite{c11}. However, in the construction of the matrix
game in our approach, the pursuers are not required to adopt the
relay pursuit strategy ($p_1$) at all times. They can adopt instead
an alternative pursuit strategy ($p_2$). Not having the pursuit
strategy fixed at all times is important in the learning process, to
increase the robustness of the evader's learned policy. The evader's
actions, denoted by $e_j \in \mathcal{A}_e, j \in \{1,..,N_e\}$, are
described as follows:
\begin{enumerate}
\item $e_1$: the evader engages in pure evasion from the nearest pursuer.
\item $e_2$: the evader performs a ``collective evasion'' maneuver which considers all the pursuers at once.
\item $e_3$: the evader heads directly towards the target.
\item $e_4$: the evader moves normal to the line of sight from the closest pursuer, that is, normal to the direction in action $e_1$.
\end{enumerate}
The physical realization of the discrete actions can be any vector
direction. The physical realization of each of these
actions is straightforward except for the action $e_2$, which is explained below. The
schematic physical realizations of the discrete actions of the
evader are illustrated in Fig. \ref{evacts}.

\begin{figure}[!htb]
\centering %
\begin{tabular}{cc}
\hspace*{-0.5cm}
 \subfloat[Action $e_2$ for $N=3$]{\label{e21}\includegraphics[width=0.24\textwidth]{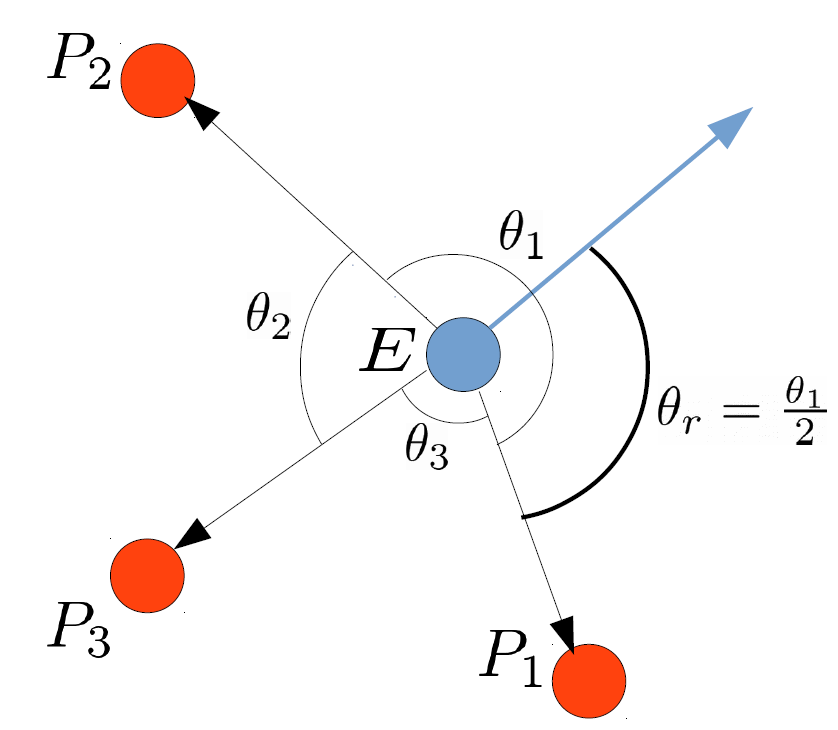}}& %0.75\linewidth
 \hspace*{-0.5cm}
 \subfloat[Action $e_2$ for $N=4$]{\label{e22}\includegraphics[width=0.25\textwidth]{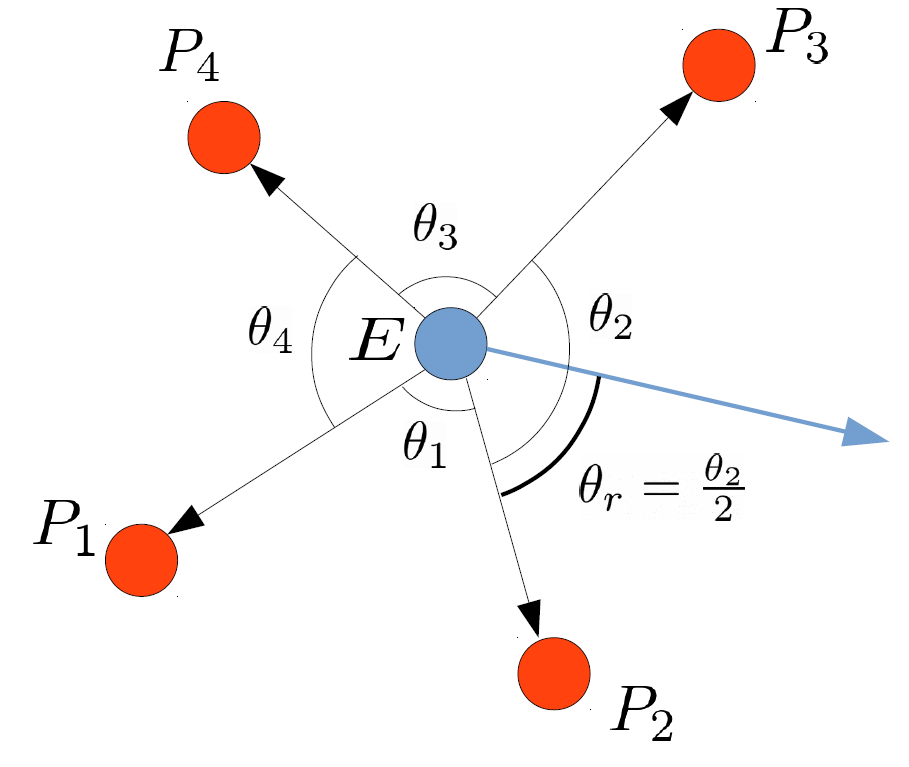}}\\  %0.75\linewidth
\end{tabular}
\caption{\small{Schematic representation of evasive action $e_2$ for
players with simple dynamics. The pursuers are represented as red
circles and the evader in blue. The blue arrow indicates the
direction of motion of the evader determined by the corresponding
evasion action.}}\label{fs:a2} \vspace{0.4 cm}
\end{figure}

\textit{Characterization of action $e_2$:} In a frame with the
evader at the origin, consider all the angles formed between the
lines of sight to two adjacent pursuers in order. The direction of
motion of the evader corresponds to the bisector of the largest
angle formed at the evader's position. This is equivalent to the
evader moving towards the largest gap between the pursuers. The
intent of this evasion action is to make the evader move away from
the entire group of pursuers at once (group evasion action), instead
of evading only one in particular. This action is useful when the
direction of optimal evasion from one pursuer (say the $i^{th}$ one)
puts the evader directly in the line of sight of another pursuer
(say the $j^{th}$ one, $j\neq i$). If $\theta_i$ is the angle
between the lines of sight from the evader to the $i^{th}$ pursuer
and to the $(i+1)^{th}$ pursuer, then we calculate the resultant
direction of motion of the evader as follows:
\begin{align*}
\bm{u}_e=\left[ \cos (\theta_r+\omega_{i_m}) ~~ \sin  (\theta_r+\omega_{i_m}) \right]^{\mathrm{T}}
\end{align*}
where $\omega_i :=\measuredangle \bm{r}_i$ denotes the polar angle
of the vector $\bm{r}_i:=\bm{x}_e-\bm{x}_i$, and
\begin{align*}
\theta_i:&=\omega_{i+1}-\omega_i, ~~\theta_i \in [0,2 \pi], \\
i_m:&=\arg \max_i \theta_i,\\
\theta_r:&=\frac{\theta_{i_m}}{2},
\end{align*}
for $i=\{1,...,N-1\}$. If $i=N$, we set $\omega_{N+1}:=\omega_1$ to
preserve the order of rotation. The evasion action is illustrated in
Fig. \ref{fs:a2}.

The evader (shown in blue) should move along the blue arrow
according to the action $e_2$. In Fig. \ref{e21}, the evader is
outside the polygon formed by the pursuers (shown in red) as
vertices, and the evader's direction of motion is between pursuers
$P_1$ and $P_2$. In Fig. \ref{e22}, the evader is inside the polygon
of the pursuers, and the preferred bisector lies between $P_2$ and
$P_3$. If two angular gaps are equal, the evader favors the angular
gap formed by the longer line of sight to a pursuer.

\subsection{Q-learning for the matrix payoffs}
The payoff matrix $M_k$ at each stage $k$ has a fixed size for any
number of pursuers, and the discrete action sets comprehensively
describe the qualitative effects of the players' actions in the
game. However, for the multi-agent game ($N>1$), the optimal payoff
function, and consequently the entries of the payoff matrix, are not
readily available analytically. Hence, we use min-max Q-learning to
construct the entries of the payoff matrix $M_k$ at each stage.

Before we describe the Q-learning procedure, we must introduce a new
state space where the Q-learning process can take place. This is
because it is not practical to carry out the learning process
effectively using position vectors as states and physical directions
as corresponding actions (this would be an intractable approach). In
the position space, a set of optimal actions learned by the players
from a particular initial configuration, does not enable them to
maneuver correctly from other initial configurations, because the
environment is not static. Further, the speeds of the players and
their number largely influence the weights learned, which in turn
influence the optimality of each action for the players.
Consequently, we are unable to extend the results (learned weights)
to other environments that are similar but with a different number
of players and with different speeds. The rewards for the
reinforcement learning procedure also cannot be explicitly
position-dependent, since the reward for the same state-action pair
in another game with different initial conditions would be
different.

In this context, we introduce a new state space in which the states
are transformed representations of the players' position vectors and
also encode other properties of the pursuit-evasion game. In
particular, we recognize that the times-to-capture and the
time-to-target are sufficient to accurately describe and decide the
progress of the game. Moreover, it is the ratio of the
time-to-target to the time-to-capture or the time to intercept
($\phi_T/\phi_c$ or $\phi_T/\phi_s$ respectively), which reflects
the safety (in the sense of staying sufficiently far away from the
pursuers) and target-reaching ability of the evader at all times.
With these considerations, we next describe the state space in which
the learning process is carried out.

\subsection{Learning state space}
The learning state space $\Psi$ is a subset of $\mathbb{R}_{\ge
0}^m$, where $\mathbb{R}_{\ge 0}^m:=\{ \bm{z} \in \mathbb{R}^m: z_j
\ge 0,~j \in \{1,...,m\}\}$. The learning state variables are
implicitly dependent on the position and velocity of the players at
a given stage. For the multi-agent PEG, $m=4$. At any stage $k \in
\{1,...,{K}\}$, let us introduce the following variables:
\begin{enumerate}
\item The least min-max time-to-capture of the evader over all the pursuers, which is given by
\begin{equation}
\psi_c(k;\bm{x}_e):=\min_i \phi_c(\bm{x}_e, \bm{x}_i).
\end{equation}
\item The least minimum time-of-intercept of the evader by a pursuer when the evader is moving
in a randomly chosen and fixed direction that is known to pursuers,
which is defined as follows
\begin{equation}
\psi_a(k;\bm{x}_e, \bm{u}_e):=\min_i \phi_a(\bm{x}_e,
\bm{x}_i,\bm{u}_e),
\end{equation}
where$\psi_a(k;\bm{x}_e, \bm{u}_e)$ corresponds to the
time-of-intercept when the evader executes action $e_2$.
\item The minimum time for the evader to reach the target state in the absence of
pursuers which is defined as follows:
\begin{equation}
\psi_T(k;\bm{x}_e,\bm{x}_T):=\phi_T(\bm{x}_e,\bm{x}_T).
\end{equation}
\item The least minimum time-of-intercept of the evader by a pursuer
when the evader heads directly towards the target location which is
given by
\begin{equation}
\psi_s(k;\bm{x}_e,\bm{x}_T):=\min_i \phi_s(\bm{x}_e, \bm{x}_i,\bm{x}_T).
\end{equation}
\end{enumerate}
For ease of notation, we denote the learning state variables as
explicit functions of the stage $k$, while in fact, they are
functions of the position vectors at time $t= k \Delta t$. Let
$\mathcal{S}:\mathbb{R} \rightarrow \left[0,1\right]$ be a sigmoid
function. At stage $k$, the state vector in the learning space $\Psi
\subseteq \mathbb{R}_{\ge 0}^m$ is denoted by
$\bm{\psi}(k;\bm{x}_e,\bm{x}_T)$ and defined as follows:
\begin{align*}
\bm{\psi}(k;\bm{x}_e,\bm{x}_T):=\Big[ \mathcal{S}
\big(\dfrac{_{\psi_c}}{^{\psi_T}}\big)~
\mathcal{S}\big(\dfrac{_{\psi_a}}{^{\psi_T}}\big)
~\mathcal{S}\big(\dfrac{_{\psi_s}}{^{\psi_T}}\big)~ \psi_T
\Big]^{\mathrm{T}}.
\end{align*}

\begin{figure}[!htb]
\centering
 \includegraphics[width=0.5\textwidth]{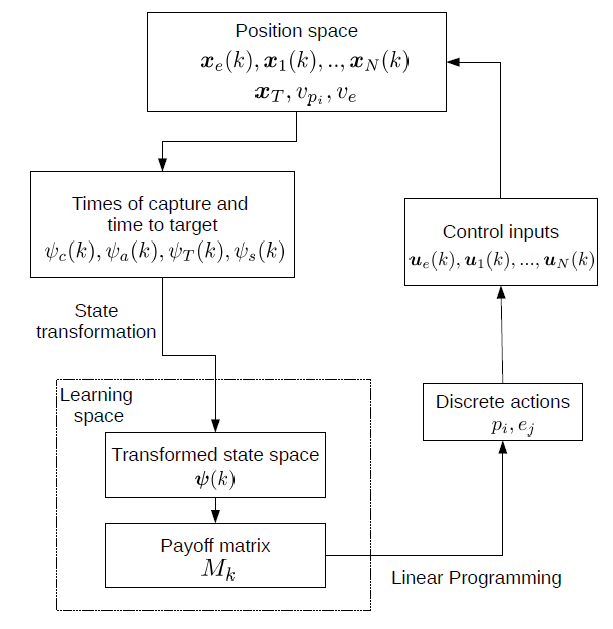}
\caption{\small{Learning applied to matrix game framework for pursuit-evasion, shown for a single stage.}}
\label{qlscheme}
\vspace{0.4 cm}
\end{figure}

The first three state variables depend on the ratios of the
time-metrics introduced in Section~\ref{mat:form1} with respect to
the time-to-target (this choice is justified by the fact that the
evader is the maximizing agent in the two-player static game). We
wish to maximize the ratio of the time-to-capture (or intercept) to
the time to reach the target. The following are the desirable
attributes of the learning state space:
\begin{enumerate}
\item The learning state space is a subset of $\mathbb{R}_{\ge 0}^m$ for all $N$.
This means that for any number of pursuers, $N$, the learning state
space has a \textit{constant dimension}. This greatly reduces the
computational expense associated with solving the matrix game at
each stage.
\item The state variables of the learning space furnish a \textit{feature-based} representation of
the original position space. In other words, we capture the essential features of the game using
just $m$ variables, instead of $2(N+1)$ position variables. The feature-based representation is
important in identifying states in the game which are far apart in the position space, but
are similar in terms of imminent capture or proximity to the target.
\end{enumerate}
Note that to propagate the dynamics forward in time and perform
state updates, we are still required to operate in the physical
space. In the next section, we will describe in detail how we use
min-max Q-learning to obtain the entries of the payoff matrix. The
training of the players in the transformed state space enables the
players in the game to maneuver correctly in environments other than
the one used for training. The schematic of the learning process
applied to the matrix game is shown in Fig. \ref{qlscheme}.

\section{Approximation of the payoff matrix with min-max Q-learning}
Let $\mathcal{T}_q:\mathbb{R}_{\ge 0}^m \times \mathcal{A}_e \times
\mathcal{A}_p \rightarrow \mathbb{R}_{\ge 0}^m$ be the state
transition function which maps the state $\bm{\psi}(k)$ at stage $k$
to the state at the next stage, $k+1$, if the actions chosen by $P$
and $E$ at stage $k$ are $p_i$ and $e_j$, respectively. We write
\begin{align*}
\bm{\psi}(k+1)=\mathcal{T}_q(\bm{\psi}(k),p_i,e_j).
\end{align*}
The transition function is determined by the dynamics of the
players. In practice, we choose to perform this state update in the
position space rather than directly in the learning space. Let the
function $\mathcal{R}_q:\mathbb{R}_{\ge 0}^m \times \mathcal{A}_p
\times \mathcal{A}_e \rightarrow \mathbb{R}$ be the reward received
by $E$ for each state transition. Further, let
$\mathcal{H}_q:\mathbb{R}_{\ge 0}^m \times \mathcal{A}_e \times
\mathbb{R}^2 \rightarrow \left[ -1,1\right]$ be the ``heading'' function. The heading function maps the state $\bm{\psi}(k)$ to the component
$\vartheta$ of the evader's
unit velocity, obtained by projecting $E$'s velocity on the line of sight to the target $x_T$.
For instance, if the action chosen by $E$ at stage $k$ is $e_j$, we
have %%
\[\mathcal{H}_q(\bm{\psi}(k),e_j,\bm{x}_e;
\bm{x}_T) := -\dfrac{_{\mathrm{d}}}{^{\mathrm{dt}}} \phi_T(\bm{x}_e(t);
\bm{x}_T) \big \rvert_{t=k \Delta t}.\]

The above relation follows directly from the definition of the
time-to-target for the evader in the absence of all pursuers, that
is, $\phi_T(\bm{x}_e(t); \bm{x}_T) :=\| \bm{x}_T  - \bm{x}_e(t) \|
/v_e $. The Q-function, denoted by $\mathcal{Q} :\mathbb{R}_{\ge
0}^m \times \mathcal{A}_p \times \mathcal{A}_e \rightarrow
\mathbb{R}$, maps every tuple (state, pursuers' action, evader's
action) to a payoff value for $E$. Given the definition of the
Q-function at stage $k$, for all $p_i$ for $ i \in \{1, \dots,
N_p\}$ and for all $e_j$, for $j \in \{1, \dots, N_e\}$, the payoff
matrix $M_k = [M_k(i,j)]$ is set to be:
\begin{equation}
M_k(i,j):=\mathcal{Q}(\bm{\psi}(k),p_i,e_j;\bm{x}_e,\bm{x}_1,
 \dots, \bm{x}_N,\bm{x}_T).
\end{equation}
Since the state space is continuous, it is not practical to use a
table of Q-values to determine the optimal state-action-action
tuples. If such a grid of states is used to generate a look-up table
for the Q-function, we must rely on interpolation to calculate the
payoff values at intermediate states. A very fine grid of discrete
states would have large computational overhead, while we might still
miss out on identifying an essential ``feature'' of a given state.
For this reason, we desire \textit{generalization}, wherein we can
operate on the qualitative abstraction of a state rather its exact
quantitative value which might not have been experienced during
learning \cite{sutbarto}. For instance, a state may be encountered
during the game, which has times to capture and intercept that were
never encountered during the training process. However, this new
state with the combination of times to capture and intercept may be
relatable to another known state with its own set of capture and
intercept times, based on the criterion that in both cases, the
evader can safely reach the target. This is one such example of
extracting the qualitative information given by a state. Following
generalization, we use an approximation of the Q-function as a
weighted linear combination of a finite number of feature-based
functions of the states and actions. In this case, the weights are
learned iteratively using gradient-descent. In our implementation,
there is no restriction on the magnitude and sign of the weights,
which are all real numbers.

\subsection{Linear approximation of the Q-function}
Motivated by the previous discussion, the basis functions for the
Q-function approximation are chosen as the coordinates of the new
state obtained from executing one action of the evader and one
action by the group of pursuers simultaneously. In particular, for
all $i \in \{1, \dots, N_p\}$ and $j \in \{1, \dots, N_e\}$,
\begin{equation}
{\mathcal{Q}(\bm{\psi}(k),p_i,e_j)= \bm{w}_q^{\mathrm{T}}
\bm{\zeta}(\bm{\psi}(k),p_i,e_j; \bm{x}_T), }
\end{equation}
where both $\mathcal{Q}(\bm{\psi}(k),p_i,e_j)$ and
$\bm{\zeta}(\bm{\psi}(k),p_i,e_j)$ are dependent on the states of
all the players and the target location ($\bm{x}_e,\bm{x}_1, \dots
,\bm{x}_N,\bm{x}_T$), and where
\begin{align*}
\bm{\zeta}(\bm{\psi}(k),p_i,e_j;\bm{x}_e;\bm{x}_T):=\begin{bmatrix}
\mathcal{H}_q(\bm{\psi}(k),e_j,\bm{x}_e; \bm{x}_T)\\ (\mathcal{T}_q(\bm{\psi}(k),p_i,e_j))_{1}
\\ (\mathcal{T}_q(\bm{\psi}(k),p_i,e_j))_{2} \\ (\mathcal{T}_q(\bm{\psi}(k),p_i,e_j))_{3}  \end{bmatrix}.
\end{align*}
%where $\bm{\psi}(k+1)=\mathcal{T}_q(\bm{\psi}(k),p_i,e_j)$ and
%$\bm{\zeta}(\bm{\psi}(k),p_i,e_j)$, $\bm{w}_q \in \mathbb{R}^{m}$.
%%
%%
Note that $\bm{w}_q$ is independent of $k$ and $\bm{\psi}(k)$, as
well as the choice of actions by the players. For a zero-sum game,
the min-max Q algorithm aims to iteratively approximate
\cite{critsurv} the Q-function $\mathcal{Q}(\cdot)$ that satisfies
the following equation:
\begin{multline}
\mathcal{Q}(\bm{\psi}(k),p_i,e_j;\bm{x}_e,\bm{x}_T)=\mathcal{R}_q(\bm{\psi}(k),p_i,e_j;\bm{x}_e,\bm{x}_T)\\
+\min_{i'}  ( \mathrm{row}_{i'}(M_{k+1}) \bm{\pi}_{e_k} ), 
\end{multline}
where $\mathrm{row}_{i'}(M_{k+1})$ denotes the row vector that corresponds to the $i'$-th row of the matrix $M_{k+1} =[M_{k+1}(i',j')]$ with $M_{k+1}(i',j') := \mathcal{Q}(\mathcal{T}_q(\bm{\psi}(k+1)),p_{i'},e_{j'};\bm{x}_e,\bm{x}_T)$.
The min-max Q-learning process is outlined in Algorithm \ref{LA}.

\begin{algorithm}

\Input{$\mathcal{A}_e$, $\mathcal{A}_p$, $\mathcal{R}_q$ and $\mathcal{T}_q$ (dynamics),
 $N_{tr}$ (maximum training episodes), $\alpha$ (learning rate), $\gamma$ (discount factor), $\mathrm{tol}$
 (convergence tolerance), $\delta \alpha$ (decay in learning rate), $\beta$ (exploration probability), $\delta \beta$ (decay in exploration).}

\BlankLine
 1. Initialize each element of $\bm{w}_q$ to a random sample from a uniform distribution in $(0,1)$\\
 2. Number of training game episodes $n=0$\\
 3. \While {$n\le N_{tr}$ \textrm{or} $\|\delta \bm{w}_q\|> \mathrm{tol}$ } {
        a. Pick random initial conditions $\bar{\bm{x}}_e$, $\bm{x}_T$, $\bar{\bm{x}}_i ~\forall i \in \mathcal{I}$\\
        b. Initialize $k=0$\\
        c. \While {game episode is not over}
            {
                i. Calculate $\bm{\psi}(k)$\\
                    ii. \For {$i=1$ \KwTo $N_p$}
                     {
                    \For {$j=1$ \KwTo $N_e$}
                    {
                        \scalebox{0.875}{$M_k(i,j)=\mathcal{Q}(\bm{\psi}(k),p_i,e_j)$}

                    }
                     }
                   iii. Find mixed strategies $\bm{\pi}_{e_k}$ and $\bm{\pi}_{p_k}$ \\
                   iv. $V(\bm{\psi}(k))=\min_{i} (\mathrm{row}_i (M_k) \bm{\pi}_{e_k})$\\
                   v.  With probability $\beta$ choose an action $\bar{e}$ for $E$ at random from set $\mathcal{A}_e$\\
                        otherwise,\\
                        Choose an action $\bar{e}$ for $E$ based on $\bm{\pi}_{e_k}$\\
            vi.  Choose an action $\bar{p}$ for $P$ based on $\bm{\pi}_{p_k}$\\
                   v.  $\Delta =\mathcal{R}_q(\bm{\psi}(k),\bar{p},\bar{e})+\gamma V(\bm{\psi}(k+1)$\\$~~~~~~~~~~~~~~~~~~~~~~~~-\mathcal{Q}(\bm{\psi}(k),\bar{p},\bar{e})$\\
                   vi.  $\delta \bm{w}_q= \alpha \cdot \Delta ~\bm{\zeta}(\bm{\psi}(k),\bar{p},\bar{e})$\\
                   vii.   $\bm{w}_q := \bm{w}_q+\delta \bm{w}_q$\\
                   viii. Perform one-step update of the state\\
                   ix.  $k:=k+1$

           }

        d. $\alpha:=\alpha\cdot \delta \alpha$ \\
        e. $\beta:=\beta-\delta \beta$

f. $n:=n+1$}

\Output{$\bm{w}_q$} \caption{Linear approximation of the Q-function
using min-max Q-learning.} \label{LA}
\end{algorithm}

Note that the assumption that the relay pursuit strategy is the
preferred strategy for the group of pursuers does restrict their
choice of action to $p_1$ exclusively. If we assume during the
learning process that the group of pursuers is required to use relay
pursuit at all states, then the problem becomes a decision making
problem for a single-agent in a dynamic environment. In that case,
the static game at every stage corresponds to a Markov Decision
Process (MDP) because there is only one available action for the
agent opposing the evader. In our proposed approach, we train the
evader to be able to handle both relay pursuit and a completely
non-cooperative pursuit strategy. This is choice is intended to
increase the robustness of the proposed learning-based solution
approach. Hence, we have a Markov Game \cite{c19} with the opposing
agent having two actions ($p_1$ and $p_2$) to choose from.

\subsection{Reward function} \label{sec:rew}
The choice of the reward function $\mathcal{R}_q$ directly
influences the learning process in terms of learning desired
behavior as well as the time taken to converge to a policy. In
particular, for problems with a desired terminal state (such as the
one considered in this paper), the shaping of the reward is
essential to drive the evader towards the target location while
avoiding the pursuers. There is positive as well as negative
reinforcement in the reward function. Since the time-to-capture
relative to the time to goal is an important parameter in
determining the evader's success, we define a reward function that
consists of the following three parts:
\begin{enumerate}
\item The target-heading reward, which is defined as:
\begin{equation}
\mathcal{H}_q(\bm{\psi}(k),e_j,\bm{x}_e,\bm{x}_T):=-\dfrac{_{\mathrm{d}}}{^{\mathrm{d}
t}} \psi_T(\bm{x}_e(t),\bm{x}_T) \big \rvert_{t=k \Delta t}.
\end{equation}
\item The state-transition reward, $\mathcal{R}_s(\bm{\psi}(k),\mathcal{T}_q(\bm{\psi}(k),p_i,e_j))$, which
is a function of the current state and the next state, and in
particular, is defined as:
\begin{equation}
\small
\mathcal{R}_s(\bm{\psi}(k),\mathcal{T}_q(\bm{\psi}(k),p_i,e_j)):=(\mathcal{T}_q(\bm{\psi}(k),p_i,e_j))_{3}-\psi_3(k).
\end{equation}
Note that we positively reward an increase in the time-to-capture
relative to the time-to-reach the target.
\item A terminal reward, $g(\bm{\psi}(k),e_j)$, which is used to positively reward the evader for reaching the target
%%\small
\begin{equation}
    g(\bm{\psi},e_j):=
    \begin{cases}
    +1, & \mathrm{if}~ \psi_4=0, \\
    -1, & \mathrm{if}~ \psi_1=\psi_2=\psi_3=0, \\
    0, & \mathrm{otherwise}.
    \end{cases}
    \label{rew}
 \end{equation}

\end{enumerate}

In practice, we can also determine the terminal reward based on the
state in the position space in order to avoid error due to finite
precision. Furthermore, the total reward associated with a specific
state transition is then given by
\begin{align}\label{totalrew}
\mathcal{R}_q(\bm{\psi}(k),p_i,e_j;\bm{x}_e,\bm{x}_T)& =\mathcal{H}_q(\bm{\psi}(k),e_j,\bm{x}_e,\bm{x}_T) \nonumber \\
&~~ + \mathcal{R}_s(\bm{\psi}(k),\mathcal{T}_q(\bm{\psi}(k),p_i,e_j)) \nonumber \\
&~~ + g(\bm{\psi}(k),e_j).
\end{align}

Note that it is important to make the training rewards as
independent as possible of the time step and the specific values of
the states encountered in the training process. For this reason, we
consider the limit with respect to the sampling period for the first
term of the reward. When one feature of the Q-function has either
too large or too small of an influence on the update of the weight
vector $\bm{w}$, that feature \textit{interferes} with the updates
due to the other features. We ensure that the range of the different
reward functions and state variables used in the learning process
are comparable to each other so that the interference is minimized
in the linear update of the weights.

\section{Numerical simulations}

In this section, we present numerical simulations that demonstrate
the performance of the evader using the improved policy obtained by
min-max Q-learning. The various parameters chosen for the training
process are: $N=3$, $v_{p_i}=v_e=1$ for all $i \in \mathcal{I}$,
$\Delta t=0.01$, $\ell=0.01$, $\epsilon=\ell$, $\gamma=0.9$,
$\alpha=0.1$, $\beta=0.9$, $\delta \alpha=0.9$, and $\delta \beta$
was chosen such that the final value of $\beta=0.01$, at the end of
the training which was performed over $N_{trn}=2000$ episodes of
games. First, we chose the speed of the evader to be slower than
that of the pursuers, and in particular, $v_e=0.9$. For training,
the initial conditions of the players and the target location were
randomly chosen from a uniform distribution within a two-dimensional
spatial grid of unit size.

The discount factor $\gamma \in [0,1]$ determines if the learning
player prioritizes immediate rewards or long-term rewards.
Typically, the discount factor is chosen to be closer to $1$,
indicating that long-term rewards are preferred. The choice of
learning rate $\alpha \in [0,1]$ is critical, because a consistently
high learning rate will make the learning process sensitive to every
input, thereby making it difficult to obtain convergence of weights.
On the other hand, if $\alpha$ is too small, the learning process
will take a long time. In general, we start with a nominal value of
$\alpha=0.1$ and decay the learning rate as more learning episodes
are covered. Finally, in the beginning of the learning process, some
randomness is introduced in the choice of actions of the learning
player to ensure exploration of the state-action space by the
player.

At the termination of the learning process, after the weights have
converged to a specified tolerance, the performance of the evader
using the strategy determined by the solution to the matrix game
with Q-learning were compared with its performance when using the
strategy determined by the solution to the $N \times (N+1)$ matrix
game, which does not involve any learning. Results obtained with the
utilization of a different strategy determined by from another
matrix game with payoff matrix of size $N_p \times N_e$, where the
payoff function was taken to be the sum of the heading reward (as described in Section
\ref{sec:rew}) and the minimum value of the learning state
variables, was also compared with the others. The package cvx
\cite{c79} as well as the built-in solver in MATLAB were used to
solve the linear programming problem.

We test the policy based on the learned payoff function with a
different number of pursuers than that used in training. It is
important to note that the size of the payoff matrix as formulated
for Q-learning is independent of the number of agents in the game.
In Table \ref{t1} we see the evader's performance for about $10^3$
episodes of the multi-player game, where $N=5$, with initial
conditions sampled from a uniform distribution within a grid of size
$10$ units. Note that the grid size is also different from the
training grid size. The column headers, which correspond to
different solution methods, are to be interpreted as follows:
\begin{enumerate}
\item M-1: Matrix game with Q-Learning with a $N_p \times N_e$ payoff matrix.
\item M-2: Matrix game without Learning with a $N \times (N+1)$ payoff matrix.
\item M-3: Matrix game without Learning with a $N_p \times N_e$ payoff matrix.
\end{enumerate}
The pursuers engage in relay pursuit at all times. From Table
\ref{t1}, it is clear that in the case of a slower evader, the three
matrix-based methods perform similarly. The success rate for the
slower evader is so low because of the number of faster pursuers.
There exist initial conditions for the players such that the evader
has no chance of success, which can be determined by analyzing the
times to capture and intercept. As $N$ increases, there is a greater
chance that the evader fails to reach the target uncaptured.

\begin{table}
\caption{Outcome of games - $v_e=0.9$, $N=5$}
\begin{center}
    \begin{tabular}{ | l | l | l | l | l |}
    \hline
    {Solution Method}  & {M-1} & {M-2} & {M-3}  \\
   \hline
    Evader Captured & $89.51 \%$ & $89.84 \%$ & $89.24 \%$  \\ \hline
    Target Reached & $10.49 \% $& $10.16 \% $&  $10.76 \%$   \\ \hline
   \end{tabular}
\end{center}
\label{t1}
\end{table}

Next, we train the evader for the case when $v_e=1$. Then, we test
the evader's performance for a different number of pursuers and a
large number of random initial conditions. In this case, we expect a
much higher rate of success for the evader, since it is as fast as
the pursuers.

\begin{figure}[!htb]
\centering %
\begin{tabular}{cc}
\hspace*{-0.5cm}
 \subfloat[M-1]{\label{f11}\includegraphics[width=0.25\textwidth]{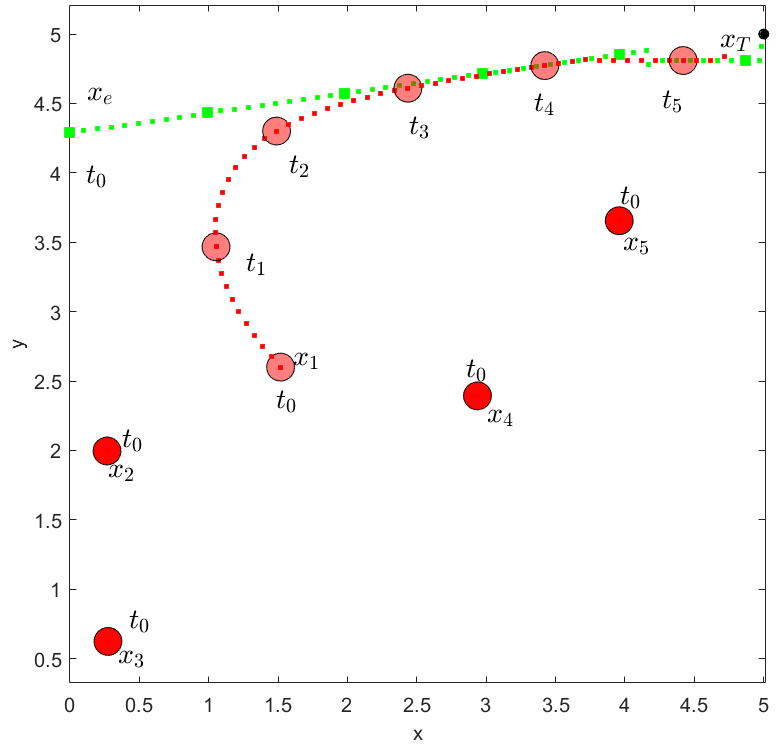}}& %0.75\linewidth
 \hspace*{-0.4cm}
 \subfloat[M-2]{\label{f12}\includegraphics[width=0.24\textwidth]{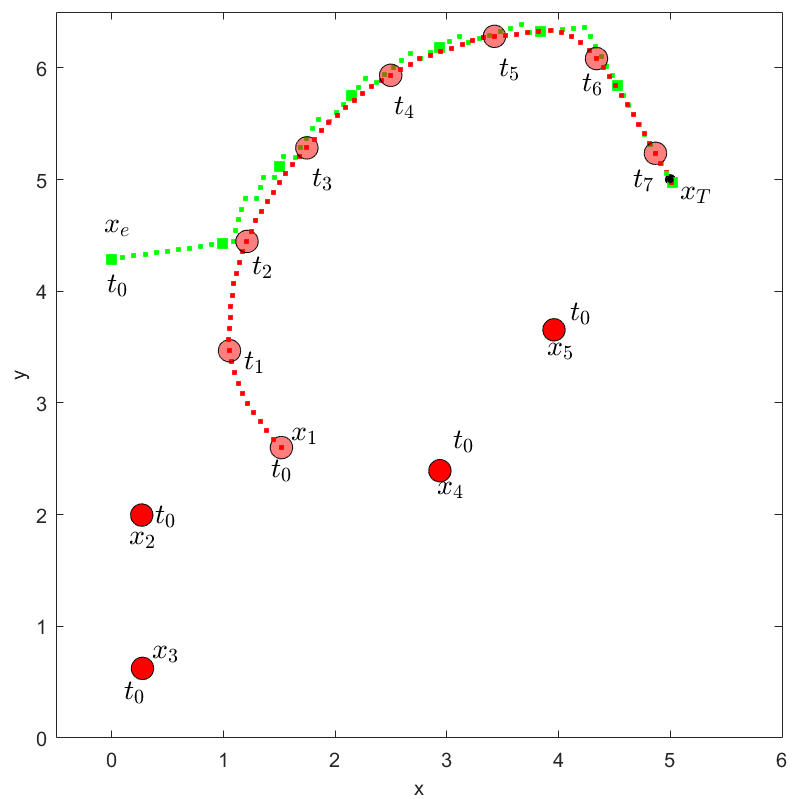}}\\  %0.75\linewidth
\end{tabular}
\caption{\small{Evolution of a pursuit-evasion game, comparing
methods M-1 and M-2, for $N=5$.}} \label{f1}
\end{figure}

\begin{figure}[!htb]
\centering %
\begin{tabular}{cc}
\hspace*{-0.5cm}
 \subfloat[M-1]{\label{f21}\includegraphics[width=0.242\textwidth]{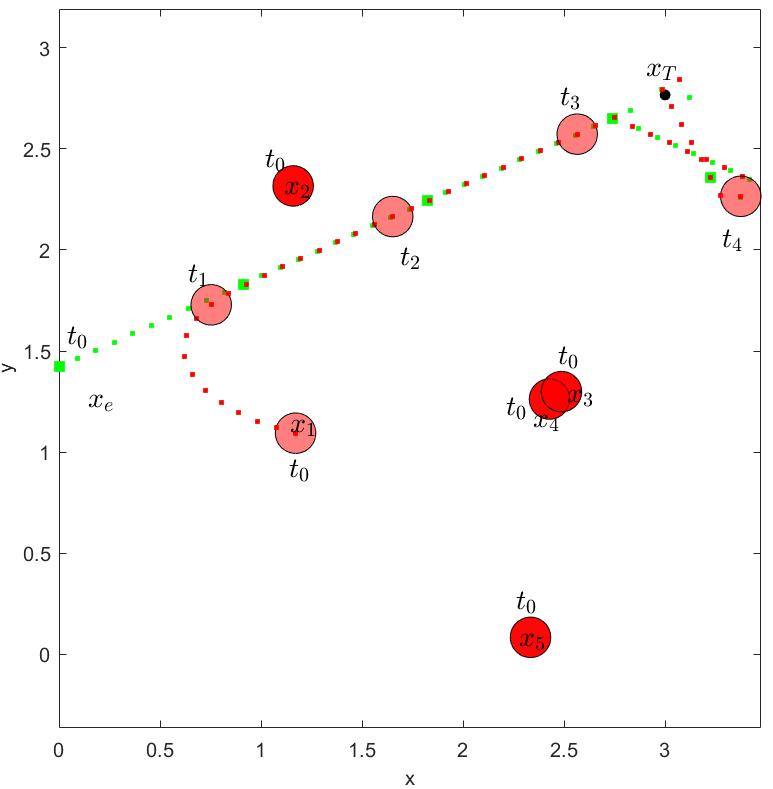}}& %0.75\linewidth
 \hspace*{-0.4cm}
 \subfloat[M-2]{\label{f22}\includegraphics[width=0.25\textwidth]{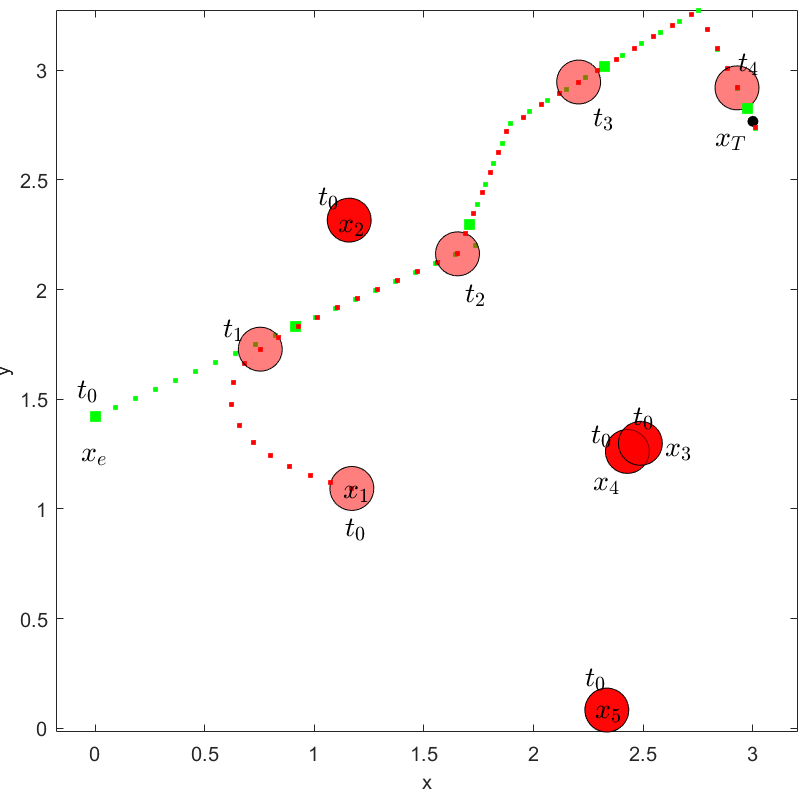}}  %0.75\linewidth
\end{tabular}
\caption{\small{Evolution of a pursuit-evasion game, comparing
methods M-1 and M-2, for $N=5$.}} \label{f2} \vspace{0.4 cm}
\end{figure}

In Table \ref{t2}, we present the results for $N=4$. We find that
all the solution approaches perform better than the previous case,
in terms of reaching the target. In terms of both avoiding capture
and reaching the target within the end time, the method using
Q-learning is better than the other two methods. Note that the
number of games that were inconclusive within our simulation time
limit is highest in the case of the evader using the learned
strategy. This means that even though the evader was unable to reach
the target, it managed to avoid capture or at least delay it until
the end of the simulation time. This is a desirable behavior for the
evader, which is not observed when the other two strategies, M-2 and
M-3, are employed. In the subsequent figures, the pursuers are
represented in red and the evader in green. The capture disks around
the pursuers and the evader (using large square markers) are shown
at certain time instances during the game. The black dot represents
the target state. The evolution of the game for two sets of initial
conditions for using the solution methods M-1 and M-2 is illustrated
in Fig.~\ref{f1} and Fig.~\ref{f2}.

\begin{table}
\caption{Outcome of games - $v_e=1$, $N=4$}
\begin{center}
    \begin{tabular}{ | l | l | l | l |}
    \hline
    {Solution Method}  & {M-1} & {M-2} & {M-3}  \\
   \hline
    Evader Captured & $2.92 \%$ & $9.55  \%$ &  $7.36 \%$  \\ \hline
    Target Reached & $87.51 \% $& $90.45 \% $&  $92.64 \%$   \\ \hline
   \end{tabular}
\end{center}
\label{t2}
\end{table}

\begin{figure}[!htb]
\centering %
\begin{tabular}{c}
\includegraphics[width=0.4\textwidth]{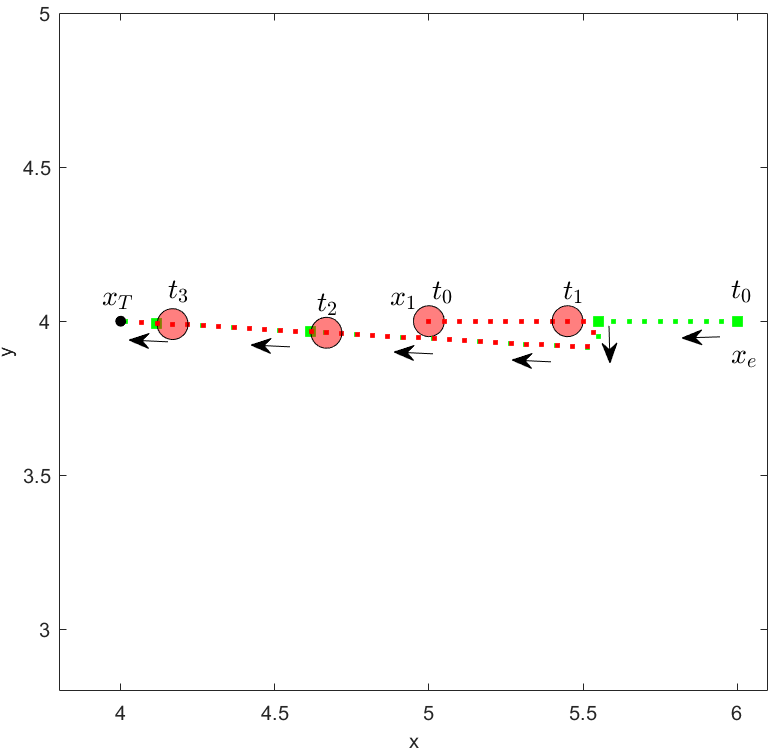}
\end{tabular}
\caption{\small{$N=1$, method M-1. The evader reaches the target in
a game where the pursuer is between the evader and the target
initially.}}
 \label{f5}
\end{figure}

In Fig. \ref{f5},
the evader uses the action $e_4$ in the normal direction, to move
towards the target, when the pursuer is initially on the evader's
line of sight to the target. The pursuer turns for a tail chase, but
the evader is able to reach the target within the required radius of
tolerance. The black arrows indicate the direction of motion of the evader.

\begin{figure*}[!htb]
\centering %
\begin{tabular}{ccc}
\hspace{-0.5 cm}
 \subfloat[$N=4$, method M-1]{\label{f31}\includegraphics[width=0.33\textwidth]{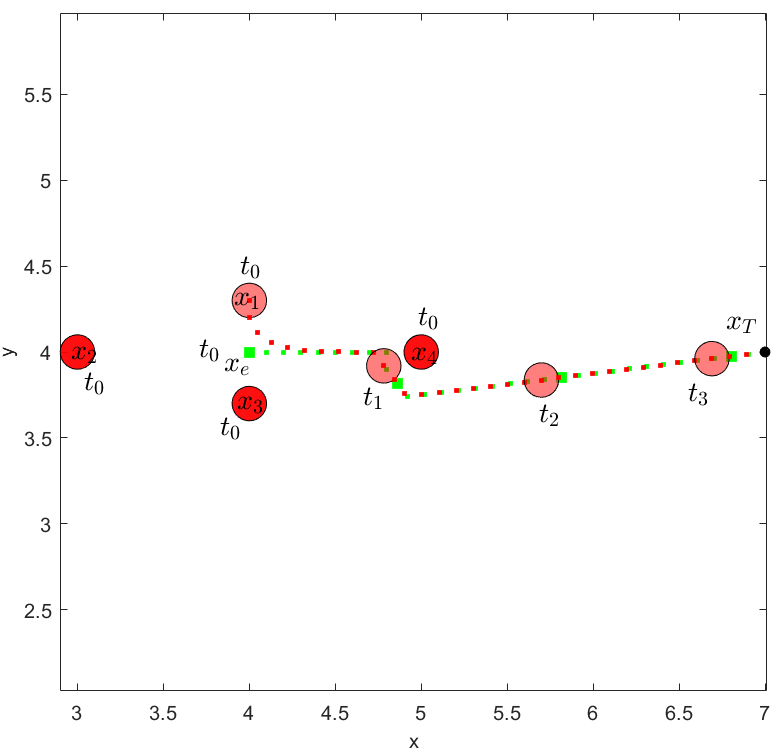}}&
 \hspace{-0.4 cm}
 \subfloat[$N=4$, method M-2]{\label{f32}\includegraphics[width=0.325\textwidth]{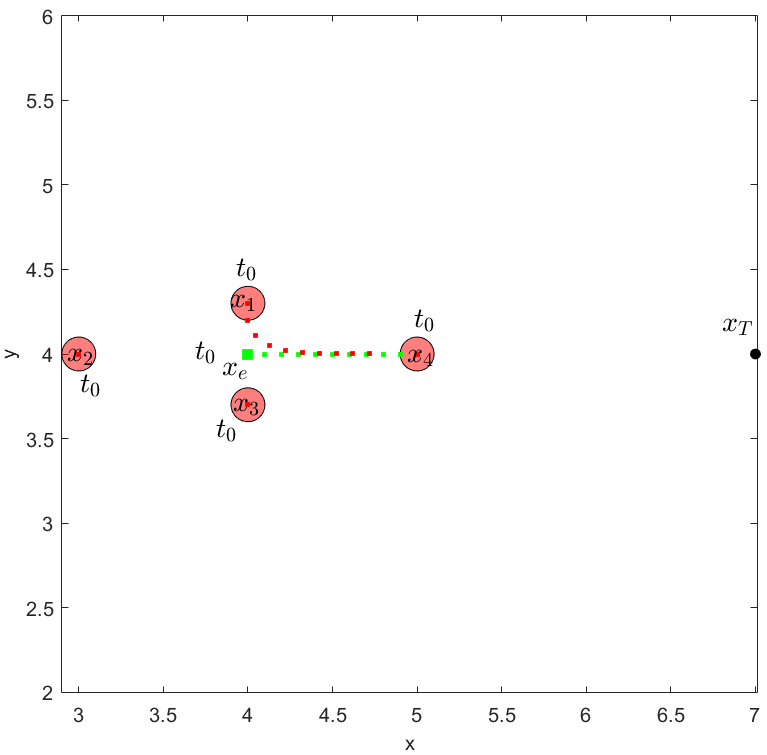}}&
 \hspace{-0.4 cm}
 \subfloat[$N=4$, method M-3]{\label{f33}\includegraphics[width=0.32\textwidth]{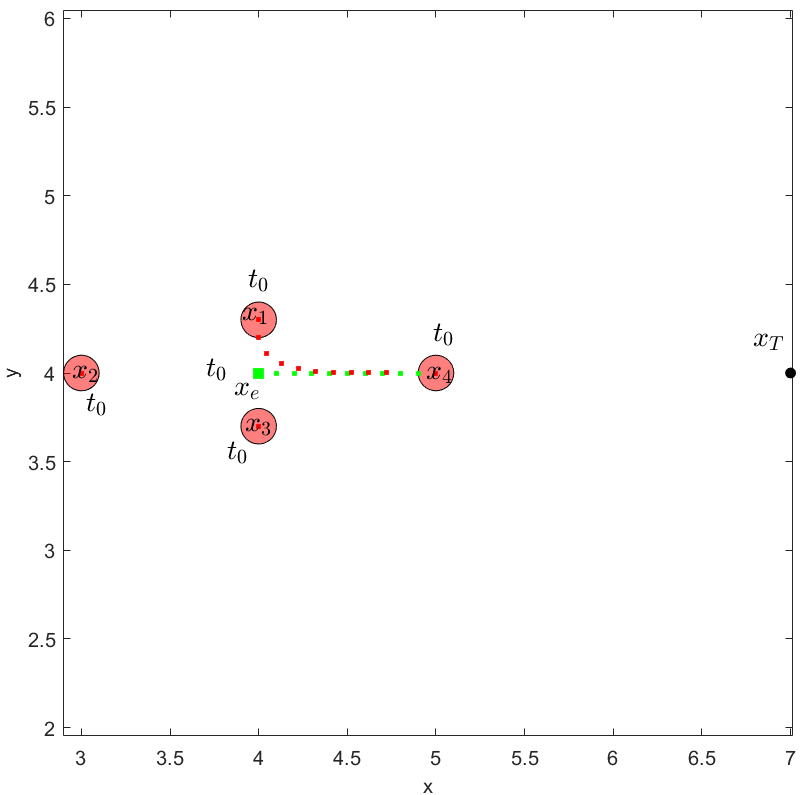}}\\
\end{tabular}
\caption{\small{Evolution of the pursuit-evasion game. The pursuers
are represented in red (with disks to show the capture radius at
certain intervals) and the evader with green square markers. The
black dot represents the target state. Note that in cases (a) and
(c) the evader reaches the target without being captured. The
players' markers are artificially enlarged and are not to scale.}}
\label{f3}
\end{figure*}

We see that even in the cases where the other methods give a good
policy for the evader, the policy determined using learned payoffs
sometimes results in the evader reaching the target quicker. This is
another advantage of learning as seen in simulations, where the
evader's decisions are not always intuitive, but lead to better
performance in terms of reaching the target. The evolution of the
pursuit-evasion game in Fig. \ref{f3} elucidates the evader's
performance when initially the evader is within the convex hull of
the pursuers. In particular, the evader's initial position is at
$[4,~4]^{\mathrm{T}}$ and the pursuers are initially at
$[5,~4]^{\mathrm{T}}, [3,~4]^{\mathrm{T}}, [4,~4.3]^{\mathrm{T}}$
and $[4,~3.7]^{\mathrm{T}}$. Clearly, the learning-based strategy
performs a swerve-like maneuver, which is not mimicked in the other
matrix payoff formulations.

\begin{figure}[!htb]
\centering %
\begin{tabular}{c}
\includegraphics[width=0.35\textwidth]{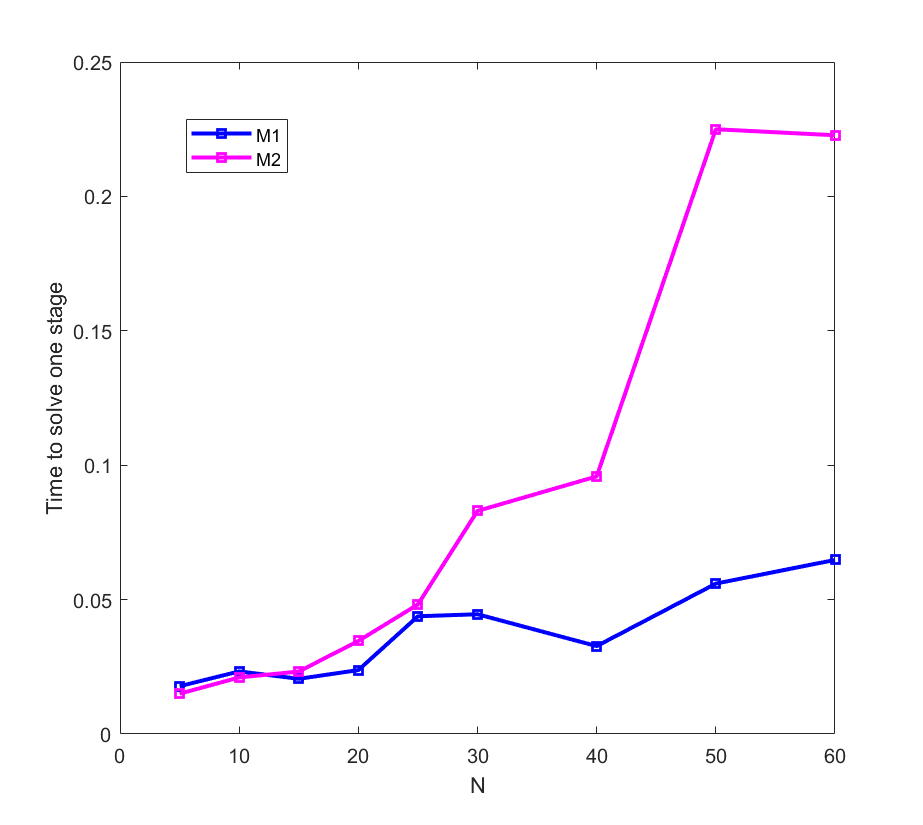}
\end{tabular}
\caption{\small{Time taken to solve a single matrix game, with the
proposed $N_p \times N_e$ matrix method M-1 in blue, and the $N
\times (N+1)$ matrix method M-2 in pink.}}\label{matcompare}
\end{figure}

Next, we examine the computational efficiency of the proposed
approach in terms of the time taken to generate the payoff matrix
and solve the matrix game for one stage. The simulations were
performed for methods M-1 and M-2 in which the payoff matrices have
different sizes. The resulting values shown in Fig. \ref{matcompare}
were derived for a grid of size $10 \times 10$, and it is seen that
the M-1 method (where the payoff matrix is of constant dimensions
$N_p \times N_e$)  is clearly advantageous in terms of computational
speed, while resulting in better performance than the M-2 method
(where the payoff matrix has size $N \times (N+1))$.

Finally, we also observe that the learning method is sensitive to
the value of the sampling period $\Delta t$ relative to the capture
radius $\ell$. While the ratio $\ell/ (v_e \Delta t)$ used during
training is maintained constant, the performance of the
learning-based method is consistent. When the ratio is changed, the
learning-based strategy executes a different sequence of moves for
the same initial conditions of the players.

\section{Concluding remarks} \label{s:concl}
In this paper, we have proposed a systematic way to compute the
evasion strategy for an evader whose goal is to reach its target
destination while avoiding capture by multiple pursuers at all
times. The proposed solution is based on a novel matrix game
formulation of the multi-agent pursuit-evasion game. The first step
of the proposed solution is to generate a dynamic payoff matrix,
whose elements are computed using a linear combination of the
time-to-target and the time-to-capture. We then refine the matrix
formulation by representing the game in a different state space. The
new state space corresponds to a feature-based representation of the game, and
the discrete actions of the players are also feature-based. The
actions chosen for each player represent the goals of that player in
the game. In the new state space and action space, we use min-max
Q-learning to learn the payoff Q-function for the zero-sum-game of
the evader against the pursuers.

One of the key results of the proposed learning-based matrix
formulation is that it is, in principle, independent of the number
of players, and their specific parameters (e.g., speed of the
pursuers). In terms of performance, the strategy that uses learning
is equally successful as the methods where the payoffs are
constructed using weighted sums. Moreover, using learning, the
evader can perform a combination of evasion maneuvers that is
difficult to explicitly generate otherwise. However, there are
dependencies of the Q-learning solution on the variables $\ell$ and
$\epsilon$ that are used during training.

We propose that
the next step is to use a deep neural network to develop an improved
non-linear approximation of the Q-function. The structure of the
learning algorithm will remain the same, however, the weight update
which is specific to the linear approximation of the Q-function will be
replaced by the corresponding weight update for a neural network
which is appropriately defined. There are some challenges pertaining
to the use of deep neural nets for Q-learning for this multi-agent
reach-avoid problem, of which two are prominent: (a) stability
concerns of a single net Q-function approximator that uses
bootstrapping, (b) convergence concerns for multi-agent Q-learning.
The former can be addressed by using a double neural network for
Q-learning, and for the latter, we can use measures such as reward
clipping and input scaling. We can also simplify the problem by
embedding the pursuers' play in the environment of the evader.
Finally, the performance of the unsupervised learning algorithm
(Q-learning in this context) heavily depends on the reward function
associated with each agent. Reward shaping, particularly for the
multi-agent reach-avoid problem under consideration, is challenging,
as it aims to determine a reward function that guarantees successful
evasion, if such evasion is feasible.

\bibliographystyle{ieeetr}

\bibliography{qlbib}

\end{document}